\documentclass[11pt]{article}

\usepackage[a4paper,margin=1in]{geometry}
\usepackage[T1]{fontenc}
\usepackage{lmodern}
\usepackage{microtype}
\usepackage{setspace}
\usepackage{textcomp}   

\usepackage{graphicx}
\graphicspath{{figures/}}
\usepackage{float}
\usepackage[font=small,labelfont=bf,labelsep=period,justification=centering]{caption}

\usepackage{booktabs}
\usepackage{array}
\usepackage{tabularx}
\usepackage{longtable}
\newcolumntype{L}[1]{>{\raggedright\arraybackslash}p{#1}}
\newcolumntype{R}{>{\raggedleft\arraybackslash}X}

\usepackage{enumitem}
\setlist{itemsep=2pt,topsep=4pt}
\usepackage{amsmath,amssymb}

\usepackage[explicit]{titlesec}
\titleformat{\section}{\Large\bfseries}{\thesection}{0.6em}{#1}
\titleformat{\subsection}{\large\bfseries}{\thesubsection}{0.5em}{#1}
\titleformat{\subsubsection}{\normalsize\bfseries}{\thesubsubsection}{0.4em}{#1}
\titlespacing*{\section}{0pt}{1.4\baselineskip}{0.6\baselineskip}
\titlespacing*{\subsection}{0pt}{1.0\baselineskip}{0.4\baselineskip}
\titlespacing*{\subsubsection}{0pt}{0.8\baselineskip}{0.3\baselineskip}

\usepackage[hidelinks,breaklinks=true,colorlinks=false]{hyperref}
\usepackage{url}
\newcommand{\us}{\,\textmu s}    
\newcommand{\x}{$\times$}        

\renewenvironment{abstract}
  {\begin{center}\bfseries\large Abstract\end{center}%
   \begin{quotation}\small\noindent\ignorespaces}
  {\end{quotation}}

\title{\bfseries Benchmarking Confidential Computing Performance\\ on NVIDIA Blackwell GPUs}
\author{%
  \begin{tabular}{c@{\hskip 1.5em}c}
    Amean Asad & Ansgar Grunseid
  \end{tabular}
  \\[1.4em]
  \href{https://confidential.ai}{Confidential.ai}
}
\date{June 2026}

\begin{document}
\maketitle

\begin{abstract}
This paper measures the performance impact of running large language model inference and training inside a Trusted Execution Environment (TEE) on NVIDIA B200 GPUs, using Intel Trust Domain Extensions (TDX) confidential VMs together with NVIDIA Confidential Computing (CC) on Blackwell GPUs. The performance impact is derived from paired confidential versus non-confidential runs on a single physical host where the only variable is the GPU CC bit and the TDX guest object in the VM launch. The main result is that confidential inference on Blackwell achieves low single-digit throughput overhead when the stack is configured correctly, at about 1-3\%. Stock inference stacks incur 30 to 40\% penalties due to avoidable configurations rather than the achievable operating point. The cost is not fully represented by a single number because it is governed by two independent axes, a fixed per-host-operation cost that amortizes as batch size grows and a per-NVLink-traffic cost that tracks the share of the step spent in encrypted collectives, and which of the two dominates is set by the workload and the software. We localize each cost to a specific encrypted boundary, give a microbenchmark that predicts the serving penalty to within a submission count, and end with concrete deployment guidance. GPU compute, energy draw, and usable memory capacity are unaffected by CC.
\end{abstract}

\section{Introduction}\label{sec:intro}

\subsection{Motivation}\label{sec:motivation}

Running a language model on infrastructure the data owner and the model owner do not control exposes both to the operator of that infrastructure. The prompt, the generated tokens, the KV cache, and the model weights are all plaintext in host memory and on the interconnect during inference, and a sufficiently privileged party, whether the hypervisor, the host operating system, another tenant, or someone with physical access, can read them. Hardware-rooted confidential computing closes this gap. Intel TDX and NVIDIA Confidential Computing mode keep data-in-use and model weights encrypted and integrity-protected the entire time they are on the CPU, on the PCIe link, in GPU memory, and on the NVLink fabric, so the operator can run the workload without being able to read or alter it.

This protection is only practically usable if the performance cost is low and bounded. A guarantee that halves throughput forces a choice between confidentiality and economics that most operators resolve against confidentiality. Therefore, the deciding question is not whether confidential inference is feasible, which it is, but how much throughput it costs, where that cost comes from, and how to drive that cost down.

This work measures the impact of NVIDIA Confidential Computing under controlled conditions, attributes it to specific hardware boundaries, identifies the mechanism that produces it, and validates the software and configuration changes that remove most of it, so that confidential inference runs almost on par with non-confidential inference.

\subsection{Problem Statement}\label{sec:problem}

The goal of this paper is to replace uncontrolled estimates of the confidential-computing inference penalty with a measured surface and an identified mechanism. Concretely, the paper aims to satisfy the following.

\begin{enumerate}
\item Every penalty is a controlled comparison, CC-on versus CC-off, on the same host, the same disks, and the same GPUs, with only the CC state changed. No penalty is inferred by comparing two differently configured runs.
\item Each cost is attributed to a specific hardware boundary that confidential computing encrypts, rather than reported as an aggregate slowdown.
\item The mechanism is validated, not just described. A synthetic microbenchmark predicts the real serving penalty from first principles, and the prediction is checked against production framework measurements.
\item The result is stated as a surface across the axes that actually move it (framework, parallelism, concurrency, sequence length), not as a single average, and the configuration that produces each number is stated alongside it.
\item The optimizations that address each identified cost are tested end to end, so that the achievable operating point is measured, not only the penalty.
\end{enumerate}

\subsection{Scope}\label{sec:scope}

This paper covers inference and training on one physical host: an Intel Xeon 6767P with eight NVIDIA Blackwell B200 GPUs, under Intel TDX and NVIDIA Confidential Computing mode. It measures throughput, latency, per-operation cost, and the NVLink bandwidth penalty across seven production models, several serving frameworks, and a range of parallelism strategies and load levels. It does not audit any security property. Whether the confidential boundary actually holds against a given adversary is out of scope. This is a performance study that takes the guarantee as given and measures its price. Attestation and confidential-VM startup overhead are not measured. Cross-node transfer, including RDMA and prefill/decode disaggregation, is out of scope for measurement. \S\ref{sec:framework} describes why it is structurally harder under CC but does not quantify it. Long-context and prefill-heavy serving beyond the ranges stated are described mechanistically where they are not measured end to end, and every such gap is flagged where it occurs and collected in the limitations of \S\ref{sec:synthesis}.

\section{Background}\label{sec:background}

\subsection{Trusted Execution Environments}\label{sec:tee}

A Trusted Execution Environment (TEE) is a hardware-enforced isolation boundary. On this stack it provides three properties the rest of the paper relies on. Confidentiality means guest memory is encrypted at the hardware level with keys the hypervisor and host operating system never possess. Integrity means tampering with that memory, by replay, corruption, or remapping, is detected by hardware. Attestation means the hardware produces a signed measurement of the software loaded at launch, chaining to the manufacturer's root of trust, so a remote party can verify what is running before trusting it.

Intel Trust Domain Extensions (TDX)~\cite{ref:intel-tdx} implements these properties for the CPU side. It places the guest VM in an encrypted, integrity-protected memory domain, a confidential VM, that the host kernel and hypervisor cannot read or alter. NVIDIA Confidential Computing (CC) mode~\cite{ref:nvidia-secureai} extends the same boundary onto the GPU. The GPU boots into a confidential state, the PCIe link between the CPU TEE and the GPU is encrypted, and on a multi-GPU system the NVLink fabric between GPUs is encrypted. A remote relying party verifies the whole arrangement through attestation: a TDX quote over the guest measurement together with an NVIDIA GPU attestation report, with the GPU releasing its CC session keys only after the attestation chain validates.

The net effect is that data-in-use, meaning activations, KV cache, prompts, and generated tokens, along with the model weights, is protected the entire time it is on the CPU, on the PCIe link, in GPU memory, and on the NVLink fabric. That protection is the guarantee whose cost the rest of this paper measures.

\subsection{The Encrypted Boundaries and Their Terms}\label{sec:boundaries}

Confidential computing on this stack encrypts four boundaries, and the paper attributes cost to each of them by name. The terms used throughout are defined here on first appearance.

\begin{itemize}
\item \textbf{PCIe bulk copies.} Host-to-device (H2D) and device-to-host (D2H) transfers, used for weight loading, inputs, and sampler readback, are encrypted with AES-GCM. NVIDIA Confidential Computing mode transfers do not use the kernel software bounce buffer (SWIOTLB). Instead the driver-managed bounce buffers are used in shared, decrypted memory plus AES-GCM over PCIe. The use of these bounce buffers creates a per-byte and per-call tax on PCIe traffic. The crypto runs in OpenSSL's \texttt{libcrypto}, reached through NVIDIA's PKCS\#11 provider. We confirmed this directly: a probe on the OpenSSL decrypt path during 8.25\,GiB of D2H traffic caught 17{,}505 calls totaling 8.44\,GiB, about 517\,KB per call, matching the bounce-buffer size.
\item \textbf{The command channel.} Every kernel submission crosses an encrypted control path to the GPU System Processor (GSP), the on-GPU microcontroller that receives host commands over an RPC channel (GSP-RPC). Under CC this channel is encrypted, so every submission pays a fixed control-path cost.
\item \textbf{NVLink encryption (NVLE).} Cross-GPU collective traffic for tensor-parallel and expert-parallel work is encrypted on the NVLink fabric. NVIDIA calls this Multi-Party Trust NVLink encryption, and we refer to it as NVLE throughout. Its cost scales with the volume of encrypted inter-GPU traffic a workload generates, so it grows with the share of wall-clock spent in cross-GPU collectives.
\item \textbf{A capability loss.} Multicast over NVSwitch (\texttt{cuMulticast}) is blocked under CC~\cite{ref:nvidia-secureai}. Kernels that depend on NVSwitch multicast, such as one FlashInfer all-reduce-plus-RMSNorm fusion, fall back to a multicast-free path.
\end{itemize}

AES hardware acceleration is in use (AES-NI, VAES, PCLMULQDQ, AVX-512). A single core tops out near 13.9\,GB/s on AES-256-GCM, and the measured CC PCIe path runs near 10\,GB/s, about 72\% of that ceiling. The shortfall is not missing hardware crypto. Each per-GPU secure session (SPDM~\cite{ref:spdm}) is single threaded in the current implementation, pinned to a single core, so host-to-device crypto does not scale across host threads.

Two latency terms recur in the serving sections. Time to first token (TTFT) is the delay from request arrival to the first generated token, dominated by prefill. Time per output token (TPOT) is the average interval between subsequent tokens, dominated by the decode loop. The paper reports both because confidential computing moves them differently.

\section{Methodology}\label{sec:methodology}

\subsection{Controlled Comparison Method}\label{sec:method-ab}

Every measurement in this paper is a paired comparison. The confidential run and the non-confidential run use the same root disk, the same data disk holding identical model bits, and the same GPU or GPUs, and differ only by the GPU CC bit and the TDX guest object in the VM launch. ``CC on'' means two things together: the per-GPU CC-mode bit set through the GPU's BAR0 register, and the guest running as a TDX confidential VM. Multi-GPU runs are sequential on identical disks, an all-eight CC-on pass followed by an all-eight CC-off pass.

\subsection{System Under Test}\label{sec:sut}

\begin{itemize}
\item Host: Intel Xeon 6767P (2 socket) with eight NVIDIA B200 GPUs on an NVLink fabric.
\item Software: Ubuntu 24.04.3, kernel 7.0.9, QEMU 11.0.0, OVMF 2025.02, TDX SEAM 2.0.14.
\item Guest GPU driver: NVIDIA 595.71.05 (open). Models in NVFP4, FP8, AWQ, and bf16.
\item Frameworks: SGLang 0.5.13.post1 and an SGLang cc-fixes branch~\cite{ref:sglang}, vLLM 0.21.0 and 0.22.0~\cite{ref:vllm}, Megatron-core 0.19.0 with TransformerEngine 2.18.0.dev0 for training.
\end{itemize}

Every configuration in the paper, with its framework, model, and parallelism, is listed in the table below. The data spans two serving frameworks across several versions and was collected over several weeks, so cross-row comparisons are directional rather than controlled. Each individual CC versus non-CC comparison within a row is controlled.

\begin{table}[H]
\centering
\small
\caption{Configurations measured in this paper, with framework, model, quantization, and parallelism.}
\begin{tabularx}{\linewidth}{@{} l L{0.20\linewidth} l L{0.16\linewidth} l X @{}}
\toprule
\textbf{Section} & \textbf{Model} & \textbf{Quant} & \textbf{Framework} & \textbf{GPUs} & \textbf{Role} \\
\midrule
5.1, 5.2 & cc\_probe (synthetic probe); Qwen3.5-0.8B/9B & n/a; FP8 & pure-CUDA; vLLM 0.22 & 1\x\,B200 & microbench, root cause \\
\midrule
5.3 & D2D + NCCL collectives & n/a & nccl-tests & 4\x\,B200 & NVLink raw cost \\
\midrule
6.1 & Nemotron-3-Super-120B-A12B & NVFP4 & SGLang 0.5.13 & 1\x, TP1 & single-GPU, overlap off \\
\midrule
6.2 & Qwen3-8B & bf16 & SGLang 0.5.13 & 1\x, TP1 & single-GPU, overlap on \\
\midrule
6.4 & Qwen2.5-72B-Instruct & AWQ-marlin & SGLang cc-fixes & 1\x, TP1 & single-GPU, patched \\
\midrule
7 & MiniMax-M2.7 228.7B/6B & NVFP4 & SGLang cc-fixes & 8\x, TP8/TP2-EP4 & multi-GPU MoE + sweeps \\
\midrule
7 & Qwen3.5-397B/17B & FP8 & SGLang cc-fixes & 8\x, DP8/EP8 & EP-over-TP control \\
\midrule
8 & in-house 41B dense; in-house \textasciitilde43B MoE & bf16/FP8 & Megatron + TE & 8\x, TP8/EP8 & training \\
\bottomrule
\end{tabularx}
\end{table}

\subsection{Measurement Constraints}\label{sec:hygiene}

There are some measurement caveats that materially affect the results we present in this paper. We present them as reproducibility notes about how to collect clean measurements under CC and not general CC constraints on a deployed workload.

\begin{itemize}
\item \textbf{Reboot between runs for clean measurements.} After a test run a CC GPU can be left in an indeterminate state by prior concurrent CUDA contexts, which reads as 100\% utilization at lower throughput and higher power. That is a stale-state artifact, not a CC cost, so the system under test should always be rebooted for clean measurements. One early uncontrolled run reported 16\% on a workload that is actually 2\% for exactly this reason. Every number here is from a clean boot.
\item \textbf{CUDA event timing is disabled by design under CC.} \texttt{cudaEventElapsedTime} can return negative values on the bounce-buffer path, because fine-grained device-side timers are restricted under CC to avoid timing side channels. This is intentional, not a defect, but it means CUDA-event timing and autotuner tactic selection cannot be trusted under CC. Timing uses the GPU \texttt{\%globaltimer} register or wall-clock throughput instead.
\item \textbf{Kernel profiling is disabled by design under CC.} NVIDIA Nsight Systems runs under CC but its CUDA and CUPTI tracing disable themselves in protected-memory mode, again by design to keep protected memory from leaking through the profiler. Host-side syscall tracing still works and shows the host-to-GSP command path as an \texttt{ioctl}, which is how the command-path cost was localized.
\item \textbf{Tracing runs are not throughput anchors.} Runs that relied on submission-level tracing were collected with GPU debug mode enabled, which is required to read per-submission timing under CC and itself adds overhead. Those runs are used to attribute where time goes, not to anchor throughput. The throughput and latency figures throughout are from separate, non-traced runs.
\end{itemize}

\section{Summary of Results}\label{sec:summary}

Confidential inference on Blackwell incurs only low single-digit overhead when the stack is configured correctly, a result NVIDIA reports independently~\cite{ref:nvidia-blog}. The headline numbers below are all from paired CC-on versus CC-off runs on one host where the only variable is the CC bit.

\begin{table}[H]
\centering
\small
\caption{The headline CC tax for well-configured confidential workloads, all from paired CC-on versus CC-off runs on one host.}
\begin{tabularx}{\linewidth}{@{} X l @{}}
\toprule
\textbf{Confidential workload (well-configured)} & \textbf{CC tax} \\
\midrule
MoE serving, 4 GPUs (TP4), decode-heavy, pinned & $\sim$1.5\% \\
MoE serving, 8 GPUs (TP8), decode-heavy, pinned & $\sim$3\% \\
Single GPU, full CUDA graphs + framework patches & under 1\% \\
8-GPU training (bf16) & 10 to 13\% \\
\bottomrule
\end{tabularx}
\end{table}

Concretely, MiniMax-M2.7 (a 229B mixture-of-experts model) served confidentially on four B200s costs about 1.5\% against non-CC. Two pinned five-repeat medians put TP8 at 2.8 and 3.6\% and TP4 at 1.5\%, and TP4 runs at 94\% of the eight-GPU throughput. The tax tracks encrypted inter-GPU traffic, so matching the tensor-parallelism degree to what the model actually requires minimizes unnecessary NVLink traffic and, with it, the encryption overhead. The high penalties, in the range of 30 to 40\%, trace to specific and avoidable configurations, either an unpatched or eager framework or a short-output prefill-heavy workload, not the achievable operating point.

The cost is not a single number because it is governed by two independent axes, and which one dominates depends on the workload and the software.

\begin{enumerate}
\item \textbf{Per-host-operation cost.} Every kernel submission and every host-device synchronization costs more under CC, because of the encrypted GSP-RPC command channel and forced-synchronous copies. This is a fixed cost per decode step, so it amortizes as batch size grows and it is removable by full CUDA graphs and framework patches.
\item \textbf{Per-NVLink-traffic cost.} Cross-GPU collective traffic is encrypted over NVLE at a roughly constant per-byte rate (\S\ref{sec:nvlink}), so a larger batch moves more encrypted bytes without raising the rate. What sets the end-to-end cost is how much inter-GPU communication the workload exposes as a share of wall-clock. Where that share is large it does not amortize with batch size and acts as a hardware floor. Where cross-GPU communication is a small fraction of the step, the NVLE contribution is correspondingly small.
\end{enumerate}

The levers that move the tax, all measured in this paper, are the following.

\begin{itemize}
\item \textbf{Use full CUDA graphs.} A single GPU in eager, overlap-on mode on shipped SGLang pays about 35 to 39\%, because CC forces the per-step token readback synchronous and serializes the scheduler. With full CUDA graphs plus the framework CC patches, the same single-GPU workload drops under 1\%. This is a framework and configuration problem, not a hardware one.
\item \textbf{Turn off overlap where the model forces it.} A single GPU with the overlap scheduler off (a Mamba hybrid, or any run with \texttt{disable\_overlap\_schedule}) pays about 2\%, flat. That is the residual hardware floor.
\item \textbf{Serve decode-heavy and long-context.} The encrypted prefill all-reduce is the dominant cost and sits on the NVLink critical path, while decode hides behind memory latency. On the production MoE the CC tax falls steeply as generation lengthens, reaching about 3.6\% at realistic output lengths and within measurement noise at 32k context.
\item \textbf{Prefer expert and data parallelism over tensor parallelism.} Removing attention tensor parallelism removes the attention NVLE traffic that otherwise grows with context.
\item \textbf{Budget for training separately.} Eight-GPU CC training costs about 10 to 13\%, entirely encrypted collective communication.
\end{itemize}

We also measured three places CC could plausibly have cost something and does not. GPU compute carries almost no CC overhead, since compute-bound prefill and GEMMs are unaffected. Energy usage is also unaffected, and in some cases lower, because the added bounce-buffer latency leaves the GPU idle more, so it draws less power. VRAM costs one fixed 700\,MB carveout on a B200, which the driver reserves at initialization for allocations it moves into protected framebuffer. It does not scale with the workload and does not move the 183\,GB usable figure, but it should be accounted for when sizing KV cache against the VRAM ceiling.

The sections that follow establish this result from the bottom up. \S\ref{sec:localization} measures the per-operation cost with a microbenchmark and localizes it on the data path. \S\ref{sec:single-gpu} shows the single-GPU tax and that it is gated by the framework's overlap behavior. \S\ref{sec:multi-gpu} measures the multi-GPU surface on the production MoE and pins the headline number. \S\ref{sec:training} covers training. Sections~\ref{sec:framework} through~\ref{sec:recommendations} give the mechanism model, deployment guidance, and limitations.

\section{Cost Localization on the Data Path}\label{sec:localization}

On this stack the confidential computing cost is a communication cost, not a compute cost. This section proves that by measuring the marginal price of a single kernel submission, assembling those submissions into a realistic decode step, and confirming the prediction on a production framework, then measuring the raw NVLink penalty that dominates once multiple GPUs are involved.

\subsection{Per-Operation PCIe Crypto Cost}\label{sec:pcie-crypto}

NVIDIA Confidential Computing transfers bypass the kernel software bounce buffer. Under CC the kernel's bounce-buffer tracepoint fires only for virtio disk and network traffic, never for the GPU, which uses its own driver-managed bounce buffers plus AES-GCM over PCIe (\S\ref{sec:boundaries}). A PCIe transfer-size sweep on a single CC GPU shows two regimes.

\begin{table}[H]
\centering
\small
\caption{PCIe transfer-size sweep on a single CC GPU. Bulk transfers of 1\,MB and up pay per-byte AES-GCM at about 0.1\,ns/byte. Small transfers are dominated by a fixed cipher setup of 3 to 6\us.}
\begin{tabular}{l r r r}
\toprule
\textbf{Transfer} & \textbf{GB/s} & \textbf{libcrypto CPU \%} & \textbf{ns/byte} \\
\midrule
4\,KB & 0.13 & 8.7\% & 0.767 \\
64\,KB & 1.86 & 9.0\% & 0.057 \\
256\,KB & 4.93 & 28.2\% & 0.068 \\
1\,MB & 7.21 & 49.6\% & 0.081 \\
16\,MB & 9.41 & 75.1\% & 0.095 \\
64\,MB & 9.64 & 76.9\% & 0.096 \\
\bottomrule
\end{tabular}
\end{table}

Bulk transfers of 1\,MB and up are per-byte AES-GCM at about 0.1\,ns/byte, roughly 10\,GB/s, with the crypto at about 77\% of one core. Small transfers of 64\,KB and below, which is the decode-sampling regime, are dominated by a fixed 3 to 6\us cipher setup that is independent of size. The crossover is around 256\,KB to 1\,MB.

That crypto does not scale across host threads. With a single CUDA context on one GPU, throughput is flat and H2D actually degrades under oversubscription.

\begin{table}[H]
\centering
\small
\caption{Confidential host-device crypto does not scale across host threads. About one core of AES-GCM runs at a time per GPU.}
\begin{tabular}{l r r}
\toprule
\textbf{Threads} & \textbf{H2D GB/s} & \textbf{D2H GB/s} \\
\midrule
1 & 10.2 & 10.7 \\
4 & 10.2 & 10.9 \\
8 & 7.8 & 10.0 \\
16 & 5.4 & 9.8 \\
\bottomrule
\end{tabular}
\end{table}

Only about one core of AES-GCM runs at a time per GPU. The way to scale confidential crypto is across GPUs, not across host threads.

\subsection{Command-Path Cost and the Serving-Penalty Model}\label{sec:cmd-path}

The single most useful diagnostic is a synthetic pure-CUDA microbenchmark, \texttt{cc\_probe}, with no Python, no PyTorch, and no serving framework. By stripping out the framework it isolates the actual CC cost and shows where it lives, free of the scheduler noise present in a real server trace. It is the experiment that explains the high serving numbers rather than merely measuring them.

First, the elementary cost, meaning the marginal price of one kernel submission and one synchronization, measured directly.

\begin{table}[H]
\centering
\small
\caption{The marginal price of one kernel submission and one synchronization, measured directly. A bare synchronization is free. The submission pays about 12\us.}
\begin{tabular}{l r r l}
\toprule
\textbf{Metric} & \textbf{CC off} & \textbf{CC on} & \textbf{CC tax} \\
\midrule
per kernel submission & 3.45\us & 15.7\us & $\sim$12\us (3.5\x) \\
sync only & 1.62\us & 1.67\us & $\sim$0 \\
\bottomrule
\end{tabular}
\end{table}

Every kernel submission costs about 12\us more under CC, which is the encrypted GSP-RPC command path to the GPU, while a bare synchronization is free. An independent measurement study of Hopper H100 CC mode reports the same control-path structure and attributes it to the GSP firewall and the SPDM session rather than to bulk crypto~\cite{ref:demystified}. The synchronization is not the problem. The submission is. In an eight-GPU CC partition with Fabric Manager up, the per-submission cost rechecks higher, at about 31\us per launch (clock-locked, not a ramp artifact), so it scales with CC-partition size, not just the CC bit. Treat 12\us as a floor.

Second, the probe assembles those submissions into a realistic decode step. \texttt{cc\_probe model} is a pure-CUDA transformer-decode skeleton (batch 8, 36 layers of 5 cuBLAS GEMMs each plus an LM head and a D2H sample per step, about 181 kernel launches per step), run graphed versus eager, CC-off versus CC-on.

\begin{table}[H]
\centering
\small
\caption{A full CUDA graph collapses the per-submission CC tax from 2.41\x to 1.04\x per decode step in the synthetic probe.}
\begin{tabular}{l r r l}
\toprule
\textbf{Mode} & \textbf{CC off \textmu s/step} & \textbf{CC on \textmu s/step} & \textbf{CC tax} \\
\midrule
eager & 2644 & 6358 & 2.41\x \\
graph (everything captured) & 2500 & 2589 & 1.04\x \\
\bottomrule
\end{tabular}
\end{table}

This one experiment establishes the full mechanism. Eager mode replays all 181 submissions on the host every step, each paying the 12\us CC tax, so CC is 2.41\x slower. Capture the entire step into one graph and the submissions vanish, so the CC tax collapses to 1.04\x. GPU compute itself carries almost no CC overhead. The model that falls out is simply:
\[
\text{CC decode tax} = (\text{host submissions per step}) \times \text{about } 12\text{\,\textmu s}
\]

A smaller decode-loop variant (8 kernels plus argmax per step) shows the same collapse in absolute terms: eager plus D2H sample at 172.6\,\textmu s/step versus graph plus D2H sample at 45.2\,\textmu s/step, of which the sampling D2H synchronization is only 5.3\us, a 3.8\x win from bundling.

The complication for real serving is that production frameworks do not capture the whole step into one graph. The 1.04\x graph figure above is an unrealistic floor. A real cudagraphed decode in vLLM is piecewise, split at every attention layer, so it still issues about 185 host submissions per step. At 12\us each that is about 2.28\,ms of pure CC overhead per step. The lever, therefore, is submission count, not crypto. Fewer graph splits and fuller capture collapse the 185 toward a handful.

A real vLLM run confirms the synthetic prediction (Qwen3.5-9B, vLLM 0.22, single B200).

\begin{table}[H]
\centering
\small
\caption{A real vLLM run confirms the synthetic prediction. The CC overhead is 2.28\,ms per step, exactly about 185 piecewise-split submissions.}
\begin{tabular}{l r r l r}
\toprule
\textbf{Config} & \textbf{CC on tok/s} & \textbf{CC off tok/s} & \textbf{CC tax} & \textbf{libcrypto CPU \%} \\
\midrule
decode, cudagraph & 1179 & 1772 & 1.50\x & 0.50\% \\
decode, eager & 127 & 246 & 1.94\x & 2.10\% \\
\bottomrule
\end{tabular}
\end{table}

Working backward from the measured numbers, CC-off at 1772\,tok/s is 4.51\,ms/step and CC-on at 1179 is 6.79\,ms/step, so the CC overhead is 2.28\,ms/step, which divided by 12\us is about 185 submissions per step, exactly the piecewise split count. The synthetic probe predicted the serving penalty to the submission. That 2.28\,ms is 50\% of the step at batch 8 and amortizes toward 15\% at production batch, since the fixed overhead is spread over a larger step. The crypto's own CPU time in decode is negligible, at 0.5\%, confirming the cost is the submissions and not the cryptographic computation. The high serving numbers are launch overhead, not bandwidth.

\subsection{Encrypted NVLink Bandwidth and Latency}\label{sec:nvlink}

Once more than one GPU is involved, a second cost appears: every byte of cross-GPU collective traffic is encrypted over NVLE, at a roughly constant rate per byte. Measured on four B200s on the same NUMA node:

\begin{table}[H]
\centering
\small
\caption{NVLE bandwidth and latency on four B200s on the same NUMA node. The per-byte rate is about 10 to 18\% of bandwidth, and small cross-GPU writes pay about 4\x latency.}
\begin{tabular}{l r r r}
\toprule
\textbf{Metric} & \textbf{CC on (NVLE)} & \textbf{CC off} & \textbf{Delta} \\
\midrule
D2D unidir read, Copy Engine & 8070\,GB/s & 9170\,GB/s & $-$12.0\% \\
D2D unidir write, Copy Engine & 8278\,GB/s & 9292\,GB/s & $-$10.9\% \\
D2D bidir write, Copy Engine & 16{,}411\,GB/s & 18{,}471\,GB/s & $-$11.2\% \\
D2D unidir read, SM-based & 7693\,GB/s & 9388\,GB/s & $-$18.0\% \\
D2D unidir write, SM-based & 7017\,GB/s & 8591\,GB/s & $-$18.3\% \\
NCCL all\_reduce & 156\,GB/s & 185\,GB/s & $-$10\% \\
NCCL all\_to\_all & 130\,GB/s & 149\,GB/s & $-$10\% \\
P2P short-write latency (median) & 14.5\us & 3.7\us & $\sim$4\x \\
\bottomrule
\end{tabular}
\end{table}

NVLE costs about 11\% of Copy Engine bandwidth, about 18\% of SM-based bandwidth, and about 10\% at the NCCL collective level, where pipelining hides per-transfer overhead. The latency hit is about 4\x for small cross-GPU writes, which is what hurts mixture-of-experts expert dispatch. These are per-unit rates, independent of batch size and of total traffic volume. Separately, CC PCIe H2D and D2H are about 15 to 19\x slower than non-CC (3.5\,GB/s versus 51.6\,GB/s H2D), but this only matters at load time, because weights and KV cache stay resident once loaded.

These two costs, the per-submission command-path cost of \S\ref{sec:cmd-path} and the constant-rate NVLE penalty here, make up the entire CC penalty on this stack. How much of the second a deployment pays is set by the share of the step spent in cross-GPU collectives, which the rest of the paper measures.

\section{Single-GPU Overhead and Overlap Gating}\label{sec:single-gpu}

On a single GPU there is no encrypted NVLink, so the only CC cost is the per-submission command-path cost of \S\ref{sec:cmd-path}. Whether that cost shows up as 2\% or 39\% turns entirely on one framework behavior: whether the overlap scheduler is active. This section shows both ends of that range on the same silicon and the same workload, then shows the framework patch that removes the expensive end.

\subsection{Overlap Off: the Residual Hardware Floor}\label{sec:overlap-off}

The Nemotron-3-Super-120B-A12B-NVFP4 model is a Mamba hybrid, which forces \texttt{disable\_\-overlap\_\-schedule\-=True} because the state update depends on the previous step. Neither arm gets overlap, and decode is fully CUDA-graphed in both. The runs use SGLang 0.5.13.post1 at TP1 with no NVLink, temperature 0, and two graph configurations in each CC state: a decode-graph-only config and a full-graph config with piecewise graphs on. Note that enabling piecewise makes SGLang drop the CUTLASS FP4 mixture-of-experts backend on this GPU and fall back to auto. Throughput was flat so the swap was benign, but it is not a perfectly clean single variable.

Output throughput (tok/s):

\begin{table}[H]
\centering
\small
\caption{Output throughput with the overlap scheduler off (Nemotron-3-Super-120B, Mamba hybrid), decode-graph versus full-graph configurations.}
\begin{tabular}{l r r r r r}
\toprule
\textbf{conc} & \textbf{non-CC dec} & \textbf{non-CC full} & \textbf{CC dec} & \textbf{CC full} & \textbf{CC tax (dec)} \\
\midrule
16 & 1195 & 1251 & 1161 & 1211 & 2.8\% \\
32 & 1930 & 1922 & 1884 & 1874 & 2.4\% \\
64 & 2637 & 2636 & 2592 & 2571 & 1.7\% \\
\bottomrule
\end{tabular}
\end{table}

Mean TTFT (ms):

\begin{table}[H]
\centering
\small
\caption{Mean time to first token. Full CUDA graph cuts TTFT by about 66\% at concurrency 16, identically in both modes.}
\begin{tabular}{l r r r r}
\toprule
\textbf{conc} & \textbf{non-CC dec} & \textbf{non-CC full} & \textbf{CC dec} & \textbf{CC full} \\
\midrule
16 & 1058 & 355 & 1114 & 380 \\
32 & 718 & 625 & 778 & 689 \\
64 & 1421 & 1286 & 1454 & 1337 \\
\bottomrule
\end{tabular}
\end{table}

Mean TPOT, power, and utilization (decode-graph config):

\begin{table}[H]
\centering
\small
\caption{Mean TPOT, power, and utilization in the decode-graph config. CC draws slightly less power at matched throughput.}
\begin{tabular}{l l l}
\toprule
\textbf{conc} & \textbf{non-CC TPOT/W/util} & \textbf{CC TPOT/W/util} \\
\midrule
16 & 12.36\,ms / 577\,W / 70\% & 12.69\,ms / 545\,W / 68\% \\
32 & 15.88\,ms / 686\,W / 80\% & 16.23\,ms / 672\,W / 79\% \\
64 & 21.81\,ms / 728\,W / 83\% & 22.17\,ms / 712\,W / 83\% \\
\bottomrule
\end{tabular}
\end{table}

Power and utilization track each other, and CC even draws slightly less power at matched throughput. Usable VRAM reads as the same 183\,GB in both modes, about 156\,GB used after load. CC reserves 700\,MB of that, where the driver moves some of its allocations into protected framebuffer, a one-time cost at initialization that does not scale with the workload. That is too small to show at whole-gigabyte precision, but it matters when comparing KV cache capacity across modes, because frameworks size the cache as a fraction of available framebuffer rather than as an absolute amount.

The decode tax is small here for three structural reasons, and only for this workload. There is no encrypted NVLink at a single GPU. Decode is fully CUDA-graphed, so the per-launch command-path crypto is already collapsed. Weights and KV cache are resident, so almost nothing crosses the encrypted PCIe path per step. This matches the per-operation model: the residual CC decode cost is a fixed roughly 11\us crypto setup, which at concurrency 64 is 11\us against a 22\,ms step, and therefore within measurement noise.

Full CUDA graph adds prefill graphing, which cuts TTFT by about 66\% at concurrency 16 (non-CC 1058 to 355\,ms, CC 1114 to 380\,ms, an absolute saving of about 703 versus 734\,ms, the same within noise), about 11 to 13\% at concurrency 32, and about 8 to 9\% at 64. It does not move decode throughput, because decode was already graphed. The saving is identical in both modes, since prefill is compute-bound and CC adds almost no overhead there. Full CUDA graph is a TTFT win, not a CC-specific one.

\subsection{Overlap On: the Lost-Overlap Penalty}\label{sec:overlap-on}

A standard dense model runs with the overlap scheduler on, on shipped SGLang 0.5.13 with no CC patch. Qwen3-8B in bf16:

\begin{table}[H]
\centering
\small
\caption{Overlap-on decode on shipped SGLang, Qwen3-8B. CC forces the per-step token readback synchronous, serializes the scheduler, and costs 34 to 39\%.}
\begin{tabular}{l r r r r r}
\toprule
\textbf{conc} & \textbf{non-CC} & \textbf{CC} & \textbf{CC tax} & \textbf{non-CC util} & \textbf{CC util} \\
\midrule
16 & 3828 & 2513 & 34.4\% & 67\% & 52\% \\
32 & 6931 & 4534 & 34.6\% & 69\% & 57\% \\
64 & 11{,}137 & 6805 & 38.9\% & 74\% & 57\% \\
\bottomrule
\end{tabular}
\end{table}

At concurrency 64, TPOT is 5.48\,ms non-CC versus 8.28\,ms CC (plus 51\%), TTFT is 259 versus 409\,ms (plus 58\%), and power is 687 versus 543\,W.

The mechanism is visible in the utilization column. With overlap on, non-CC holds 74\% utilization by hiding the device-to-host token readback behind compute. Under CC that readback goes through the bounce buffers and is forced synchronous, so the scheduler serializes. This is the bounce buffer tax in its most expensive form. The GPU stalls each step, utilization drops to 57\%, throughput drops about 39\%, and power drops because the GPU is idle. The inverted power signature, where CC uses less power, is lost overlap, not a hardware fault.

\begin{figure}[H]
\centering
\includegraphics[width=0.92\linewidth]{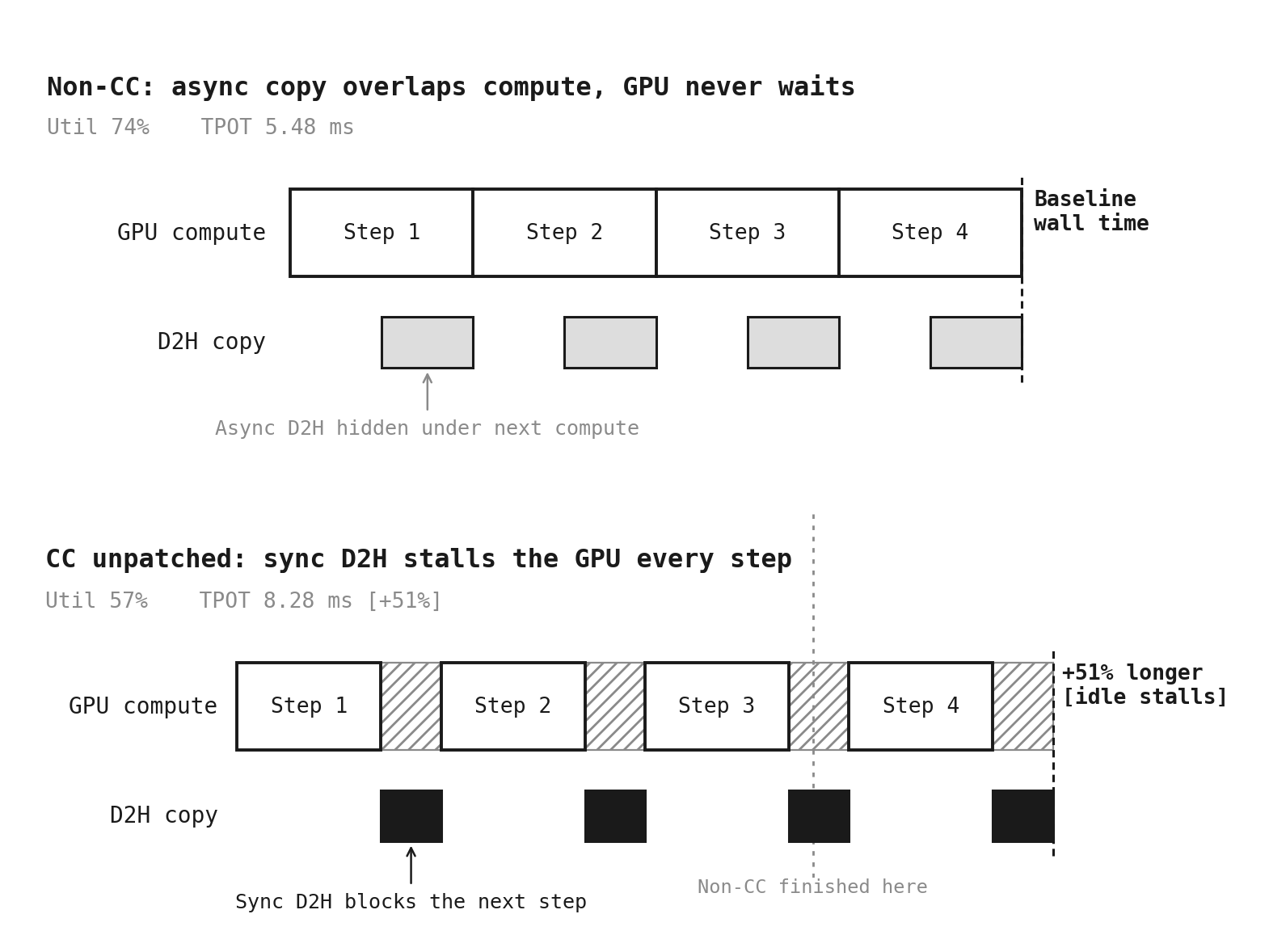}
\caption{Under CC the per-step D2H token readback is forced synchronous and stalls the GPU between steps, whereas non-CC hides it behind the next step's compute. The async-D2H-worker change of \S\ref{sec:framework} restores the hidden-readback behavior.}
\label{fig:overlap}
\end{figure}

\subsection{The Single-GPU Range, Summarized}\label{sec:single-gpu-range}

\begin{table}[H]
\centering
\small
\caption{The single-GPU range, summarized. Same silicon, same CC mechanism, two very different taxes set by one framework setting.}
\begin{tabular}{l l l l}
\toprule
\textbf{Regime} & \textbf{Model} & \textbf{Overlap} & \textbf{CC tax} \\
\midrule
overlap off & Nemotron-120B (Mamba) & forced off & $\sim$2\% \\
overlap on & Qwen3-8B (dense) & on & $\sim$35 to 39\% \\
\bottomrule
\end{tabular}
\end{table}

Same silicon, same CC mechanism, two very different taxes set by one framework setting. The 2\% figure is real but holds only when overlap is already off, and it is not the general single-GPU CC serving tax. Note the relationship to \S\ref{sec:cmd-path}: the fixed per-submission cost amortizes with batch, which is why vLLM projects toward 15\%, but the lost-overlap cost on an overlap-enabled framework does not amortize on its own and stays at about 35\% until the framework is patched, which \S\ref{sec:framework} covers.

\subsection{The Framework Patches Remove the Penalty}\label{sec:patched-single}

Applying the NVIDIA CC framework patches (an asynchronous D2H worker, plus the SGLang cc-fixes branch) to a standard single-GPU server erases the overlap-on penalty almost entirely. Qwen2.5-72B-AWQ at TP1 on one B200, CC patched versus non-CC:

\begin{table}[H]
\centering
\small
\caption{The framework patches erase the overlap-on penalty almost entirely (Qwen2.5-72B-AWQ, TP1, CC patched versus non-CC).}
\begin{tabularx}{\linewidth}{@{} X l @{}}
\toprule
\textbf{Sweep} & \textbf{CC penalty} \\
\midrule
concurrency 1024-in / 256-out, c1 to c32 & $-$0.2\% to $+$0.6\% (within noise) \\
10 input/output-shape configs at c32 & median 1.2\%, worst 6.5\% (4096-in / 1024-out) \\
\bottomrule
\end{tabularx}
\end{table}

The concurrency sweep is flat at essentially zero. The asynchronous D2H worker restores the compute/copy overlap that \S\ref{sec:overlap-on} showed CC otherwise destroys. The only configs that still cost a few percent are long-input, long-output shapes, where the larger per-step host-device traffic leaves a small residual. With a single GPU there is no NVLink, so once the per-step axis is patched out nothing remains. This is the single-GPU half of the patched story. \S\ref{sec:multi-gpu} is the multi-GPU half, where the NVLink floor does remain.

\section{Multi-GPU Overhead and Parallelism Dependence}\label{sec:multi-gpu}

On multiple GPUs the CC cost is no longer in the computation and no longer only in the command path. It is determined by how much encrypted cross-GPU traffic the parallelism generates and where that traffic sits relative to compute. Two patterns bound the range. Pipeline parallelism and expert or data parallelism move activations point-to-point or route experts, generating little synchronous collective traffic, and cost essentially nothing under CC. Tensor parallelism does an all-reduce on the full activation every layer every step, and that traffic is encrypted over NVLE, which is the expensive pattern. The tax is set by which pattern dominates and by how much of wall-clock is the communication-bound prefill all-reduce versus memory-bound decode.

The rest of this section quantifies that on the production model, MiniMax-M2.7 (a 228.7B mixture-of-experts model with 6B active parameters) on eight B200s, and arrives at the headline operating point, namely that a properly configured confidential MoE serve costs low single digits. The pieces are the two cost axes (\S\ref{sec:two-axis}), the framework patches (\S\ref{sec:patch-recovery}), the parallelism lever (\S\ref{sec:parallelism-lever}), and the full production surface with the pinned number (\S\ref{sec:tp8-surface}).

\subsection{The Two-Axis Cost Model}\label{sec:two-axis}

The two-axis model is the organizing idea for everything below. Both axes are stated in one consistent unit, the added share of step time under CC. The per-host-operation cost, meaning the per-step submission and synchronization overhead of \S\ref{sec:cmd-path}, is a fixed amount of host time per step. Its share falls as batch grows, because the step lengthens while the added time stays constant, and full CUDA graphs plus the framework patches remove it within measurement noise (\S\ref{sec:patched-single}). The per-NVLink-traffic cost, the NVLE CC tax, is the constant per-byte NVLE rate of \S\ref{sec:nvlink} applied to the share of the step spent in encrypted cross-GPU collectives. A larger batch moves more encrypted bytes but does not raise the per-byte rate. What rises with concurrency on decode-heavy serving is the exposed share itself, which saturates once compute and communication grow together, giving the measured plateau of about 5\% in \S\ref{sec:tp8-surface}. That plateau is a level, not a ceiling. The end-to-end NVLE cost is bounded by the per-byte rate itself, because the exposed share cannot exceed the whole step, and pure-collective microbenchmarks already pay the full rate (\S\ref{sec:nvlink}). That share grows with prefill intensity and tensor-parallel context (\S\ref{sec:parallelism-lever}), and where it is large it does not amortize, so it acts as a workload-dependent hardware floor. The two axes are independent and additive, each with its own lever. Removing host calls changes nothing about the encrypted bytes per token, and reducing cross-GPU traffic changes nothing about the call count. What a deployment pays is set by two independent choices, the software configuration for the per-host-operation axis and the parallelism and workload shape for the NVLE axis.

Both axes are visible directly on the production MoE. In the MiniMax-M2.7 concurrency sweep (\S\ref{sec:tp8-surface}) the per-step axis is already gone at low concurrency, since patches plus full graphs put the tax at roughly zero from concurrency 1 to 8, and the NVLE traffic share then fills in as concurrency rises, plateauing in the low single digits for decode-heavy traffic. On this stack the all-reduce runs as a custom all-reduce that works under CC (the IPC handshake passes and the algorithm is unchanged), and the only CC-specific difference is the multicast-free IPC fusion workspace, not a fallback to NCCL.

\subsection{CUDA Graph Capture Requirement}\label{sec:graph-required}

The synthetic probe of \S\ref{sec:cmd-path} shows that eager decode pays the roughly 12\us per-submission CC tax on every one of the roughly 180 launches per step, while a full graph collapses them to a single replay. On a multi-GPU MoE the un-graphed launch count is even larger, so the gap between eager and graphed widens further under CC than the already-large gap on non-CC. Full CUDA graph capture is not optional for confidential serving.

\subsection{Patch Recovery of the Per-Step Axis}\label{sec:patch-recovery}

The NVIDIA CC framework patches (an asynchronous D2H copy worker, the all-reduce-plus-RMSNorm fusion ungating, and a globaltimer-based autotuner, on the SGLang cc-fixes branch) target the per-host-operation axis. The asynchronous worker moves the per-step token readback off the critical path so that CC's forced-synchronous copy no longer serializes the scheduler. On a single GPU this restores decode TPOT from 8.68\,ms back to 5.63\,ms, matching non-CC. On a single GPU this accounts for the entire penalty, and the tax goes under 1\% (\S\ref{sec:patched-single}).

On the multi-GPU MoE the patches recover that same per-step axis, and what remains is only the NVLE CC tax on the all-reduce. A utilization signature confirms that the residual is NVLE rather than idle time. Where the patched-CC throughput trails non-CC, the patched-CC GPU utilization is actually higher, meaning the GPU is busy encrypting NVLink traffic rather than stalled. How large that residual is depends entirely on the workload. It is concentrated in the communication-bound prefill all-reduce, so it is small for realistic decode-heavy serving and grows only for short, prefill-heavy traffic. \S\ref{sec:tp8-surface} measures that residual across the full surface and pins it.

\subsection{Parallelism Strategy and the Long-Context Penalty}\label{sec:parallelism-lever}

Two patched-SGLang runs, identical methodology (full CUDA graphs, overlap off, FP8/NVFP4 mixture-of-experts runner, all-reduce fusion), sweep input length at concurrency 32 with 256-token output. Only the parallelism topology differs. This is the patch-isolation setup of \S\ref{sec:patch-recovery}, with overlap and piecewise graphs disabled and short output, not the production-default config of \S\ref{sec:tp8-surface}, so the absolute penalties here run higher than that section's for that reason and are meant for the within-row topology comparison, not for cross-comparison with the surface.

MiniMax-M2.7-NVFP4, TP2 / EP4 / DP2 on 4 GPUs (attention tensor-parallel = 2):

\begin{table}[H]
\centering
\small
\caption{MiniMax-M2.7 with attention tensor parallelism (TP2). The CC penalty grows with input length, because longer prefill means more encrypted attention all-reduce traffic.}
\begin{tabular}{l r r r}
\toprule
\textbf{Input} & \textbf{non-CC} & \textbf{CC patched} & \textbf{CC penalty} \\
\midrule
4k & 1088.6 & 990.3 & 9.0\% \\
8k & 890.9 & 778.8 & 12.6\% \\
16k & 698.5 & 618.7 & 11.4\% \\
32k & 453.7 & 387.8 & 14.5\% \\
\bottomrule
\end{tabular}
\end{table}

Qwen3.5-397B-A17B (FP8), no tensor parallelism on 8 GPUs (attention TP=1, full DP-attention, experts EP=8):

\begin{table}[H]
\centering
\small
\caption{Qwen3.5-397B with no attention tensor parallelism. The CC penalty shrinks with input length, because there is no attention all-reduce to encrypt.}
\begin{tabular}{l r r r}
\toprule
\textbf{Input} & \textbf{non-CC} & \textbf{CC patched} & \textbf{CC penalty} \\
\midrule
4k & 866.7 & 770.5 & 11.1\% \\
8k & 681.8 & 625.1 & 8.3\% \\
16k & 374.3 & 366.1 & 2.2\% \\
32k & 195.4 & 191.8 & 1.9\% \\
\bottomrule
\end{tabular}
\end{table}

The two trends run opposite, and that is the point. With attention tensor parallelism (the TP2 run), the CC penalty grows with input length, from 9 to 14.5\%, because longer prefill means more encrypted attention all-reduce traffic, which is the NVLE traffic axis. With no tensor parallelism (the 397B DP8/EP8 run), the CC penalty shrinks with input length, from 11 down to 2\%, because there is no attention all-reduce to encrypt, so the only CC cost is a fixed per-step and expert-communication overhead that amortizes as per-step compute grows with context. At 32k the no-TP 397B confidential tax is about 2\%, essentially free.

The actionable lever for long-context confidential serving is therefore to prefer DP-attention plus expert parallelism over tensor parallelism, which eliminates the attention NVLE traffic that otherwise grows with context.

\subsection{The Production-Default TP8 Surface}\label{sec:tp8-surface}

Sections~\ref{sec:patch-recovery} and~\ref{sec:parallelism-lever} disabled overlap scheduling and piecewise CUDA graphs to isolate the patches. This subsection instead uses the production-default SGLang config that we recommend running: overlap scheduling and piecewise CUDA graphs both left on, plus three always-on wins (fused parallel QK-norm, the FP8/NVFP4 mixture-of-experts runner, and all-reduce fusion). MiniMax-M2.7-NVFP4 at TP8 on eight B200s, CC versus non-CC on the identical disk and workload. Every point below is a single pass over random data except the canonical 1024/2048/c32 point, which is pinned to a five-repeat-per-arm median. Single passes carry roughly plus or minus 2 percentage points of run-to-run noise, visible in the pinned-versus-single-run gap at concurrency 32, so these tables are reliable for the shape of the surface, not for any one cell to better than a couple of points.

\begin{figure}[H]
\centering
\includegraphics[width=0.7\linewidth]{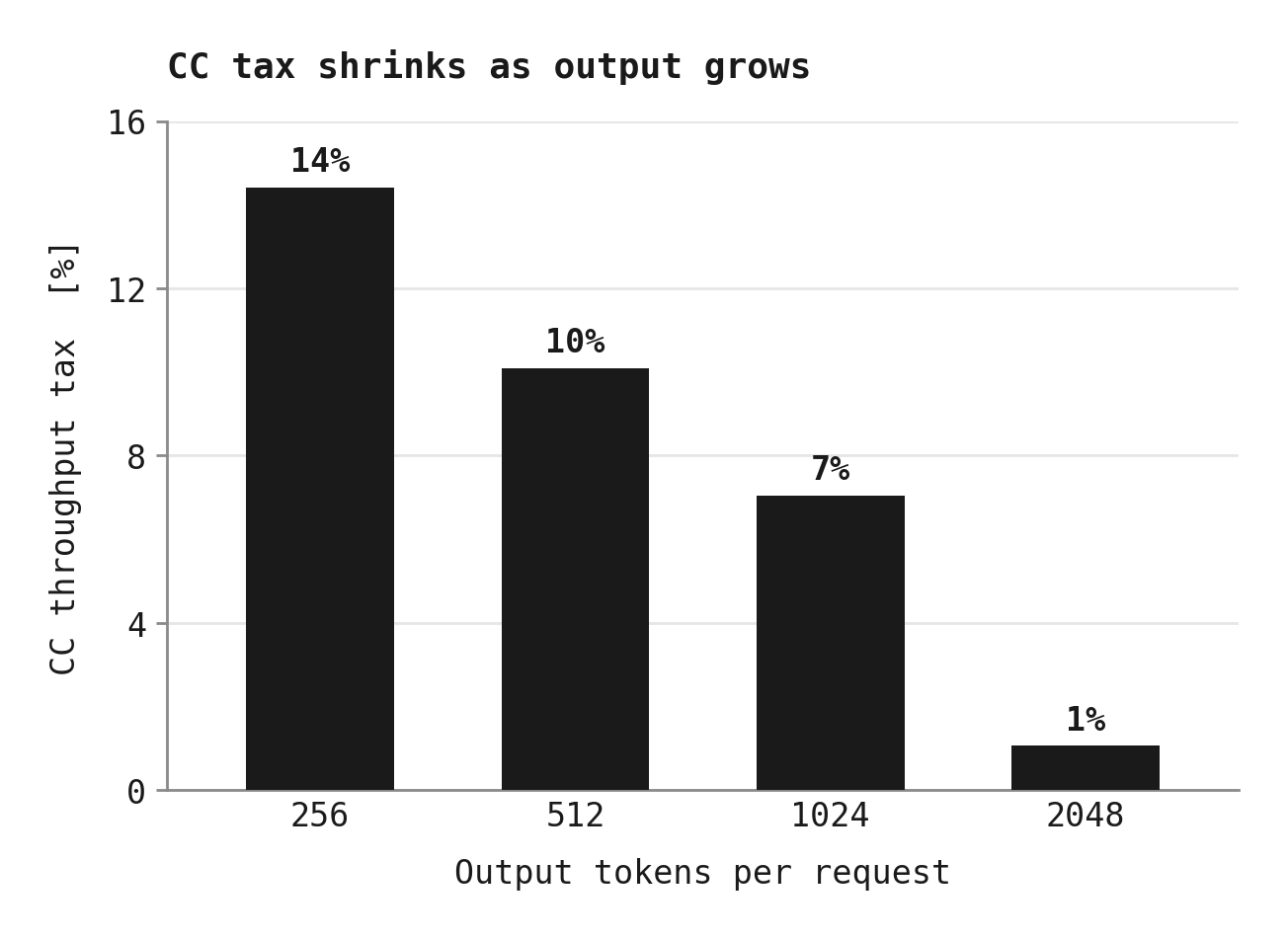}
\caption{The CC tax falls from about 14\% at short output (256 tokens) to about 1\% at realistic output (2048), as the encrypted prefill all-reduce amortizes over more decode. Production-default config, 1024-in, concurrency 32.}
\label{fig:tax-output}
\end{figure}

\begin{figure}[H]
\centering
\includegraphics[width=0.78\linewidth]{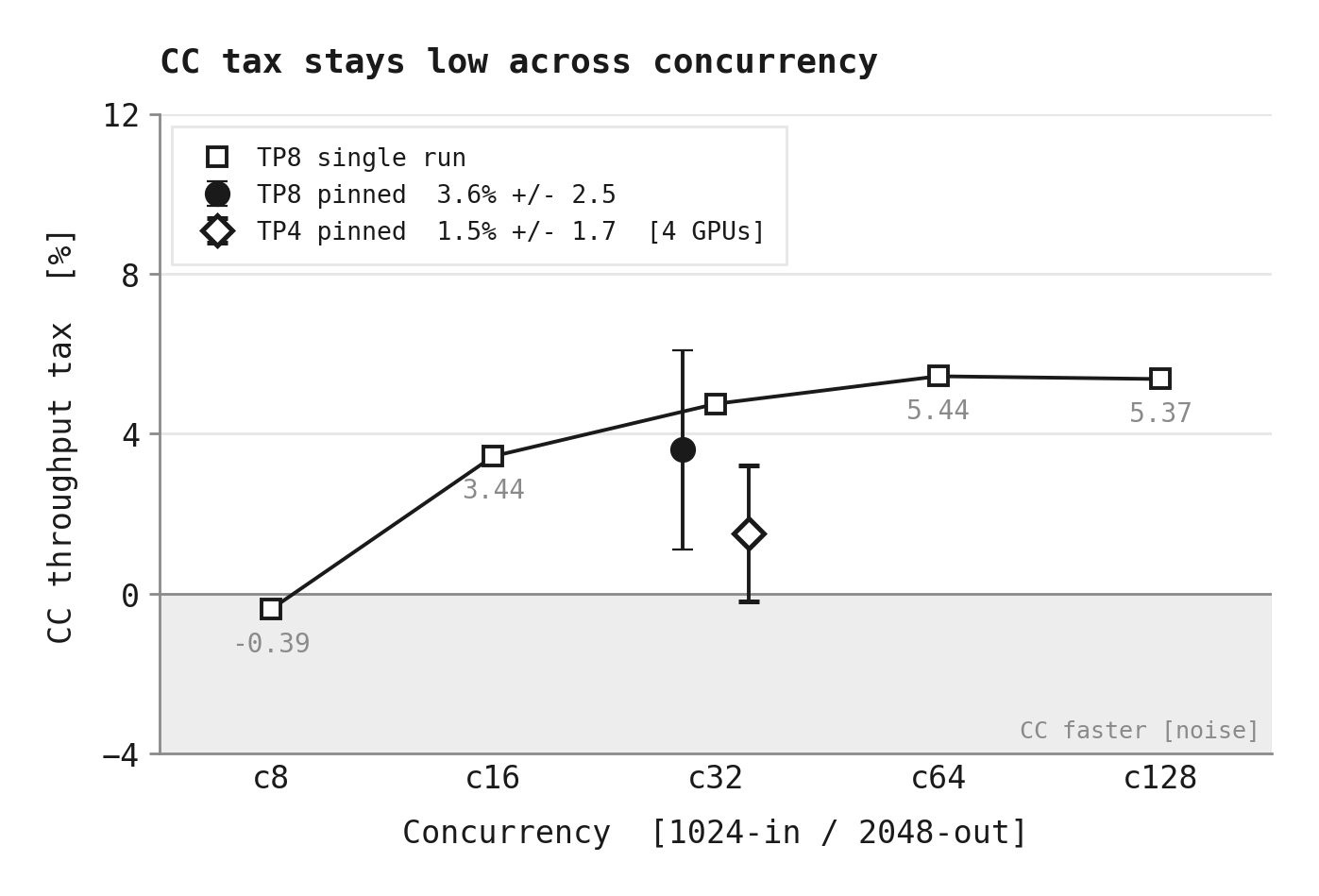}
\caption{Across concurrency at the realistic 1024-in / 2048-out shape the tax stays in the single digits and plateaus near 5\%. The canonical point at concurrency 32 is pinned by five repeats per arm at 3.6\% on TP8 and 1.5\% on TP4, the latter at 94\% of TP8 throughput. TP4 puts half as many ranks on each encrypted all-reduce, so matching the tensor-parallelism degree to what the model requires lowers the encrypted-NVLink tax at almost no throughput cost.}
\label{fig:tax-conc}
\end{figure}

Sweep one, concurrency at fixed 1024-in / 2048-out:

\begin{table}[H]
\centering
\small
\caption{Concurrency sweep at fixed 1024-in / 2048-out, production-default config. The penalty rises from noise to a plateau of about 5\%.}
\begin{tabular}{l r r r}
\toprule
\textbf{conc} & \textbf{non-CC out tok/s} & \textbf{CC out tok/s} & \textbf{CC penalty} \\
\midrule
1 & 172.7 & 176.7 & $-$2.4\% \\
8 & 892.1 & 895.6 & $-$0.4\% \\
16 & 1702.6 & 1644.0 & 3.4\% \\
32 & 2571.5 & 2449.3 & 4.8\% \\
64 & 4169.6 & 3942.6 & 5.4\% \\
128 & 6041.6 & 5717.2 & 5.4\% \\
\bottomrule
\end{tabular}
\end{table}

Sweep two, sequence length at fixed concurrency 32:

\begin{table}[H]
\centering
\small
\caption{Sequence-length sweep at fixed concurrency 32. Longer input amortizes the fixed per-step cost.}
\begin{tabular}{l r r r}
\toprule
\textbf{in / out} & \textbf{non-CC out tok/s} & \textbf{CC out tok/s} & \textbf{CC penalty} \\
\midrule
512 / 512 & 2170.9 & 2009.9 & 7.4\% \\
1024 / 1024 & 2522.6 & 2392.1 & 5.2\% \\
2048 / 2048 & 2593.2 & 2481.5 & 4.3\% \\
4096 / 1024 & 2412.8 & 2319.5 & 3.9\% \\
\bottomrule
\end{tabular}
\end{table}

Long context, concurrency 16:

\begin{table}[H]
\centering
\small
\caption{Long context at 32k input is free within measurement noise.}
\begin{tabular}{l r r r}
\toprule
\textbf{in / out} & \textbf{non-CC out tok/s} & \textbf{CC out tok/s} & \textbf{CC penalty} \\
\midrule
32768 / 2048 & 1154.6 & 1159.3 & $-$0.4\% \\
\bottomrule
\end{tabular}
\end{table}

The surface shows the two-axis model along both dimensions at once. Along concurrency the penalty rises from noise (about 0\%, even slightly negative, at concurrency 1 to 8) to a plateau near 5.4\% by concurrency 64, the NVLE CC tax filling in as all-reduce volume grows with batch. Along sequence length the penalty falls from 7.4\% at 512-in / 512-out to about 4\% at 4k input and to noise at 32k, the fixed per-step CC overhead amortizing as per-step compute grows. Across the realistic-shape surface every point stays in the single digits, and long context at 32k is free within measurement noise. The corner that reaches the low teens is short output, isolated in the output-length sweep below, where the single prefill all-reduce is not yet amortized over enough decode.

The output-length effect is the dominant driver and isolates cleanly. Under the same production-default config, fixing input at 1024 and concurrency at 32 and varying only output length:

\begin{table}[H]
\centering
\small
\caption{Output-length sweep at fixed 1024-in and concurrency 32. The single encrypted prefill all-reduce amortizes over more decode, from 14.4\% down to about 1\%.}
\begin{tabular}{l r r r}
\toprule
\textbf{output} & \textbf{non-CC out tok/s} & \textbf{CC out tok/s} & \textbf{CC penalty} \\
\midrule
256 & 1701.0 & 1456.0 & 14.4\% \\
512 & 2300.5 & 2068.4 & 10.1\% \\
1024 & 2551.4 & 2371.6 & 7.0\% \\
2048 & 2530.7 & 2504.0 & 1.1\% \\
\bottomrule
\end{tabular}
\end{table}

The trend is monotonic, from 14.4\% down to about 1\%. The dominant cost is the encrypted prefill all-reduce, which sits on the NVLink-bandwidth-bound critical path. The decode all-reduce is one token per step and hides behind memory latency, so it barely pays the tax. More output means more low-cost decode steps amortizing the single expensive prefill, so the penalty falls as generation lengthens. The \S\ref{sec:parallelism-lever} patch-isolation methodology, with overlap and piecewise graphs disabled and 256-token output, shows the same shape shifted higher in absolute terms. Under the production-default config here the short-output corner is about 14\% rather than the low twenties.

The canonical point, pinned. Single passes over random data vary by about 2 percentage points, so the canonical 1024-in / 2048-out / c32 point was run five times per arm in two independent sessions. Session one gave non-CC median 2586, CC median 2493, penalty 3.6\%. Session two gave non-CC median 2582, CC median 2510, penalty 2.8\%. The settled TP8 figure is therefore about 3\%, from 2.8 to 3.6 across two pins, with per-arm standard deviation under 2\%. What is stable either way is the shape of the surface: small, decreasing with output length and context, low single digits for realistic generation, into the teens only for short prefill-heavy traffic.

Right-sizing tensor parallelism, lower tax, near-equal throughput. Re-running the entire production-default surface at TP4 instead of TP8 puts the pinned canonical penalty at 1.5\%, about half the TP8 tax. That figure is a five-repeat median per arm, non-CC 2458 against CC 2420, with near-zero variance across the CC repeats. The shape of the surface is identical: concurrency runs about $-$3 to 4.5\%, output length falls from 15\% at 256-out to 1\% at 2048-out, and 32k context is free within noise. The mechanism is structural. TP4 puts half as many ranks on each encrypted all-reduce, so it moves less traffic over NVLE and the total NVLE cost is smaller. And it costs almost nothing in throughput. At the canonical point TP4 delivers 94\% of TP8's tokens per second (2428 versus 2571), because this MoE with 6B active parameters is not bandwidth-bound enough at these batch sizes for the extra four GPUs to add much. For confidential MoE serving the lever is concrete: match the tensor-parallelism degree to what the model actually requires rather than over-sharding it, which minimizes unnecessary encrypted all-reduce traffic and, with it, the NVLE tax, at almost no throughput cost. The insight is about the traffic, not the GPU count: a larger deployment that keeps the same tensor-parallel width per all-reduce would not pay more.

\section{Training}\label{sec:training}

Eight-GPU training with Megatron-core plus TransformerEngine, on the same overlay disk CC versus non-CC:

\begin{table}[H]
\centering
\small
\caption{Eight-GPU training with Megatron-core plus TransformerEngine. The tax is entirely encrypted collective communication. GEMM and HBM are unaffected.}
\begin{tabularx}{\linewidth}{@{} L{0.19\linewidth} l X X l @{}}
\toprule
\textbf{Config} & \textbf{Model} & \textbf{CC} & \textbf{non-CC} & \textbf{CC tax} \\
\midrule
dense bf16, TP8 & 41B dense & 15.8\,s, 550\,TFLOPs, 920\,W & 13.95\,s, 627\,TFLOPs, 960\,W & 1.13\x (13\%) \\
dense delayed-FP8, TP8 & 41B dense & 18.0\,s, 480\,TFLOPs & 14.5\,s, 600\,TFLOPs & 1.24\x \\
tuned MoE, EP8 & \textasciitilde43B MoE & 2.40\,s, 520\,TFLOPs & 2.16\,s, 575\,TFLOPs & 1.11\x (11\%) \\
\bottomrule
\end{tabularx}
\end{table}

The training CC tax is about 10 to 13\% and is entirely encrypted collective communication, meaning the tensor-parallel all-reduce and the mixture-of-experts all-to-all. GEMM and HBM are unaffected. Utilization is identical, and non-CC actually draws more power because the GPUs stall less on communication and spend that time computing. Two practical notes follow. FP8 loses extra under CC (1.24\x versus 1.04\x non-CC) because the FP8-immune collective traffic becomes a larger share of a shorter step once encryption adds overhead, so bf16 is the right default for confidential eight-GPU training. And this is the training-side restatement of \S\ref{sec:multi-gpu}: the cost lives on NVLink, not in the compute.

\section{The Framework Component of the Tax}\label{sec:framework}

Most of the single-GPU overlap loss is a framework software problem rather than a hardware one, and the fix is the same everywhere. The per-step device-to-host copy has to move onto a worker thread so that CC's forced-synchronous copy no longer serializes the scheduler. This change is not yet standard in the shipping serving frameworks, so we built and tested it ourselves against the frameworks we measured, and the same work is beginning to appear as in-development patches upstream.

We applied three changes and measured each on this stack. The first is the asynchronous D2H worker described above, which moves the per-step token readback off the critical path. The second skips host-memory pinning entirely. CC guests do not support pinned host buffers on this stack, only the driver-managed bounce buffers of \S\ref{sec:boundaries}, so requesting pinned memory provides no benefit. The third replaces CUDA-event timing in the kernel autotuner with the GPU globaltimer, because CUDA-event timing is disabled by design under CC to avoid timing side channels (\S\ref{sec:hygiene}), and otherwise the autotuner selects kernel tactics off unusable measurements. On the multi-GPU MoE we additionally ungated the all-reduce-plus-RMSNorm fusion onto a multicast-free workspace, since the multicast path is blocked under CC (\S\ref{sec:boundaries}).

The lack of pinned host buffers reaches beyond single-node serving. GPUDirect RDMA is blocked outright once GPU CC is enabled. A cross-node transfer cannot DMA directly between the NIC and GPU memory and has to bounce GPU to CPU to CPU to GPU instead. That bounce path cannot use pinned host buffers either, so it lands on the same encrypted bounce-buffer channel as ordinary PCIe traffic (\S\ref{sec:boundaries} and \S\ref{sec:pcie-crypto}) rather than the pinned-memory fast path RDMA implementations normally rely on.

Prefill/decode disaggregation depends on fast KV-cache transfer between separate prefill and decode instances, usually over RDMA. With RDMA blocked and pinned buffers unavailable, deploying prefill/decode disaggregation under CC is a harder integration problem than on a non-CC host. We did not measure end-to-end disaggregated serving under CC, so this is a gap rather than a quantified cost.

Looking forward, the bounce-buffer detour that forces this is an artifact of today's device-assignment model, not a permanent constraint. The emerging TEE Device Interface Security Protocol (TDISP) standardizes securely assigning a PCIe device such as a GPU or a NIC directly into the confidential guest, with the device's DMA traffic encrypted and integrity-protected end to end rather than routed through host bounce buffers. On a platform that supports TDISP for the GPU and the NIC together, GPUDirect RDMA and pinned-buffer fast paths become available inside the trust boundary, which would remove the bounce path this section describes and shrink the bounce buffer tax at its source. We flag this as a direction rather than a measured result, since it depends on hardware and driver support that is not yet shipping on this stack.

With these applied, the single-GPU tax drops from the 35 to 39\% of the unpatched overlap-on case (\S\ref{sec:overlap-on}) to under 1\% (\S\ref{sec:patched-single}). The unpatched penalty we measured is consistent with what the in-development upstream efforts report for the same missing async-copy behavior, in the range of 40 to 87\% at high concurrency on other models. A framework without this change shows the high tax by default. The effect on unmodified vLLM, for instance, was about 50\% at batch 8, the same unpatched signature.

The implication is that the single-GPU CC tax a team observes is as much a property of the framework version as of the hardware. The NVLE component in \S\ref{sec:multi-gpu} is the closer-to-hardware floor that software cannot remove. On the multi-GPU MoE the same changes recover the per-step axis just as they do on a single GPU, which drops under 1\%. The residual is the encrypted prefill all-reduce, which \S\ref{sec:tp8-surface} shows is small for realistic decode-heavy serving.

\section{Synthesis of the Compute Tradeoff}\label{sec:synthesis}

\begin{table}[H]
\centering
\small
\caption{The CC tax by scenario, its mechanism, and whether software can recover it. Single-GPU rows are decode-graphed. The MoE rows are the production-default decode-heavy config at the pinned 1024-in / 2048-out point, the TP8 figure pinned across two five-repeat sessions (2.8 and 3.6\%), TP4 keeping 94\% of TP8 throughput, and the EP / DP-attention row at 32k context.}
\begin{tabularx}{\linewidth}{@{} L{0.29\linewidth} L{0.17\linewidth} >{\raggedright\arraybackslash}X L{0.19\linewidth} @{}}
\toprule
\textbf{Scenario} & \textbf{CC tax} & \textbf{Mechanism} & \textbf{Recoverable?} \\
\midrule
1 GPU, overlap off (Mamba) & $\sim$2\% & residual per-sync setup & near floor \\
\midrule
1 GPU, overlap on, unpatched & $\sim$35 to 39\% & lost compute/copy overlap & yes, async D2H \\
\midrule
1 GPU, graphed + patched & under 1\% & per-step axis removed & at floor \\
\midrule
MoE TP8 & $\sim$3\%, pinned & prefill all-reduce on NVLE & no, but small \\
\midrule
MoE TP4 & $\sim$1.5\% & less NVLE traffic & no, but smaller \\
\midrule
MoE EP / DP-attention & $\sim$2\% & no attention all-reduce & near zero \\
\midrule
8-GPU training (TP/EP) & $\sim$10 to 13\% & encrypted collectives & no, hardware floor \\
\bottomrule
\end{tabularx}
\end{table}

Three places CC could plausibly have cost something, and does not.

\begin{itemize}
\item VRAM. CC reserves 700\,MB on a B200 for driver allocations moved into protected framebuffer. The reservation is fixed, so it does not move the measured 183\,GB usable figure, but it should be accounted for when sizing KV cache against the VRAM ceiling.
\item Energy. CC does not increase power draw. When it is slower it is because the GPU is idle, so it draws less power. The tradeoff is throughput and latency, not power.
\item Compute. CC adds almost no overhead for compute-bound work, meaning prefill, GEMMs, and training compute. The tax lives on the host-device and GPU-GPU boundaries.
\end{itemize}

The conclusion is that on Blackwell, confidential inference and training add almost no overhead to GPU compute, and the cost is paid on the host-device and GPU-GPU boundaries. The single-GPU price is mostly a software-architecture problem being fixed upstream. The multi-GPU price is mostly encrypted NVLink, a cost that scales with cross-GPU traffic volume and that batching does not erase.

\section{Deployment Recommendations}\label{sec:recommendations}

Distilled from the findings above, for running confidential inference on Blackwell at the lowest CC tax.

\begin{itemize}
\item Use full CUDA graphs, always. The per-submission CC tax of about 12\us each dominates decode, and graphs collapse the submissions. This is the single most effective setting and is good practice independent of CC.
\item Use a framework with the CC patches, meaning the asynchronous D2H worker. On a patched stack the single-GPU CC tax is under 1\%, while on a shipped or unpatched stack the same workload pays 30 to 40\% from lost overlap. Check for the asynchronous D2H worker before quoting any single-GPU CC number.
\item Keep weights and KV cache resident on the GPU. The encrypted PCIe path is about 10\,GB/s and does not scale across host threads. Avoid CC-mode KV offload, CPU expert offload, and weight streaming on the hot path, since they all hit the slow crypto channel.
\item Right-size tensor parallelism to what the model requires. The tax tracks encrypted inter-GPU traffic, so what lowers it is reducing that traffic, not reducing GPU count as such. Each tensor-parallel rank adds an activation to the encrypted all-reduce, so a narrower tensor-parallel width means a smaller NVLE traffic tax. The measured canonical point pays about 3\% at TP8 but about 1.5\% at TP4, at 94\% of the eight-GPU throughput, because for this MoE the extra tensor-parallel width barely adds tokens per second while it doubles the ranks on each all-reduce. Prefer expert and data parallelism over tensor parallelism for the same reason, since it removes the attention NVLE traffic. Over-sharding a model past what it needs is the failure mode to avoid.
\item Serve decode-heavy, and prefer longer outputs and longer context. The dominant cost is the encrypted prefill all-reduce, while decode hides behind memory latency, so it barely pays. On the measured MoE the CC tax falls from about 14\% at 256-token output to about 1\% at 2048, and to noise at 32k context. Short, bursty, prefill-heavy traffic is the one corner that reaches the low teens.
\item Budget the residual NVLink floor, but know it is small when configured right. Fully patched, with full graphs and decode-heavy traffic, the encrypted-NVLink tax on a multi-GPU MoE is low single digits, about 1.5 to 3\% across TP4 and TP8 at the canonical point, rising toward roughly 5\% only at very high concurrency. This part is hardware, not software, but it is far smaller than the unpatched or prefill-heavy corners.
\item Expect to profile host-side only. GPU kernel profiling and CUDA event timing are disabled by design under CC, to keep protected memory from leaking through side channels (\S\ref{sec:hygiene}), so plan observability around guest CPU-clock profiling, host-side NVTX ranges, and end-to-end throughput.
\end{itemize}

In absolute terms the tax is manageable. The measured CC configs here still serve hundreds to thousands of tokens per second per deployment, so confidential inference is production-viable. The question is how many percentage points of throughput the guarantee costs, not whether it is feasible.

\end{document}